%% file: main.tex
\documentclass[sigconf]{acmart}
\AtBeginDocument{%
  \providecommand\BibTeX{{%
    \normalfont B\kern-0.5em{\scshape i\kern-0.25em b}\kern-0.8em\TeX}}}

\usepackage{CJKutf8}
\usepackage{graphicx}
\usepackage{multirow} 

\newcommand{\add}[1]{{\textcolor{black}{ #1}}}
\newcommand{\delete}[1]{}

\copyrightyear{2024}
\acmYear{2024}
\setcopyright{acmlicensed}\acmConference[CHI '24]{Proceedings of the CHI
Conference on Human Factors in Computing Systems}{May 11--16,
2024}{Honolulu, HI, USA}
\acmBooktitle{Proceedings of the CHI Conference on Human Factors in
Computing Systems (CHI '24), May 11--16, 2024, Honolulu, HI, USA}
\acmDOI{10.1145/3613904.3642860}
\acmISBN{979-8-4007-0330-0/24/05}

\begin{document}

\title[Understanding Promises and Challenges of Online Synchronized Voice-Based Mobile Dating]{Seeking Soulmate via Voice: Understanding Promises and Challenges of Online Synchronized Voice-Based Mobile Dating}


\author{Chenxinran Shen}
\affiliation{%
  \institution{University of British Columbia}
  \city{Vancouver}
  \country{Canada}}
\email{elise.shen007@gmail.com}

\author{Yan Xu}
\affiliation{
  \institution{University College Dublin}
  \city{Dublin}
  \country{Ireland}
}
\email{yen.hsu@outlook.com}

\author{Ray LC}
\affiliation{%
  \institution{City University of Hong Kong}
  \city{Hong Kong}
  \country{Hong Kong SAR}
}
\email{lc@raylc.org}

\author{Zhicong Lu}
\authornote{Corresponding author.}
\affiliation{
  \institution{City University of Hong Kong}
  \city{Hong Kong}
  \country{Hong Kong SAR}
}
\email{zhicong.lu@cityu.edu.hk}

\begin{abstract}
\input{abstract}
\end{abstract}




\maketitle


\input{00-intro}
\input{01-related-work}

\input{02-method}
\input{03-findings}

\input{04-discussion}
\input{05-Design_Implication}

\input{06-conclusion}


\balance
\bibliographystyle{ACM-Reference-Format}
\bibliography{sample-base}










\end{document}

%% file: abstract.tex

Online dating has become a popular way for individuals to connect with potential romantic partners. Many dating apps use personal profiles that include a headshot and self-description, allowing users to present themselves and search for compatible matches. However, this traditional model often has limitations. In this study, we explore a non-traditional voice-based dating app called ``Soul''. Unlike traditional platforms that rely heavily on profile information, Soul facilitates user interactions through voice-based communication. We conducted semi-structured interviews with 18 dedicated Soul users to investigate how they engage with the platform and perceive themselves and others in this unique dating environment. Our findings indicate that \add{the role of voice as a moderator influences impression management and shapes perceptions between the sender and the receiver of the voice. Additionally, the synchronous voice-based and community-based dating model offers benefits to users in the Chinese cultural context. }
\delete{Soul users tend to form perceptions of potential partners that encompass emotions, reactions, and personalities through voice-based interactions. Furthermore, synchronous communication via voice significantly enhances the online dating experience and fosters deeper mutual affection. }
Our study contributes to \add{understanding the affordances introduced by voice-based interactions in online dating in China.}

%% file: 00-intro.tex
\section{Introduction}
The widespread adoption of Mobile Dating Applications (MDAs) has coincided with the widespread use of smartphones. A study reveals that 38\% of adults in the United States have tried MDAs at least once, highlighting the popularity of online dating as a method for people to meet potential romantic partners \cite{smith2013online}. As more and more people join these platforms, the online dating industry has attracted substantial investments and has firmly established itself as a mature market \cite{prweb, Masden2015onlinedating}.
While HCI research has been exploring the affordances of online dating sites that heavily rely on text-based communication~\cite{ingram2019looking},
voice-based online dating represents a relatively underexplored domain within the research landscape. Understanding the intricate mechanisms governing synchronous voice-based interactions in the context of online dating has received limited attention to date. By examining interactions mediated by voice, our study contributes to bridging this research gap, offering valuable insights that can benefit both researchers and dating platforms alike. These insights have the potential to drive improvements in the overall quality of communication within the realm of online dating \cite{voicedating}.
We selected Soul as the primary voice-based dating platform for our study due to its popularity as one of the biggest voice-based dating platforms globally. As of January 2021, Soul had amassed 30 million monthly active users, ranking among the top apps in its category in China~\cite{BaiduBaike}. 
We conducted in-depth interviews with 18 dedicated Soul users and analyzed the data using thematic analysis \cite{thematic}. 

\begin{figure}
    \includegraphics[width=8cm]{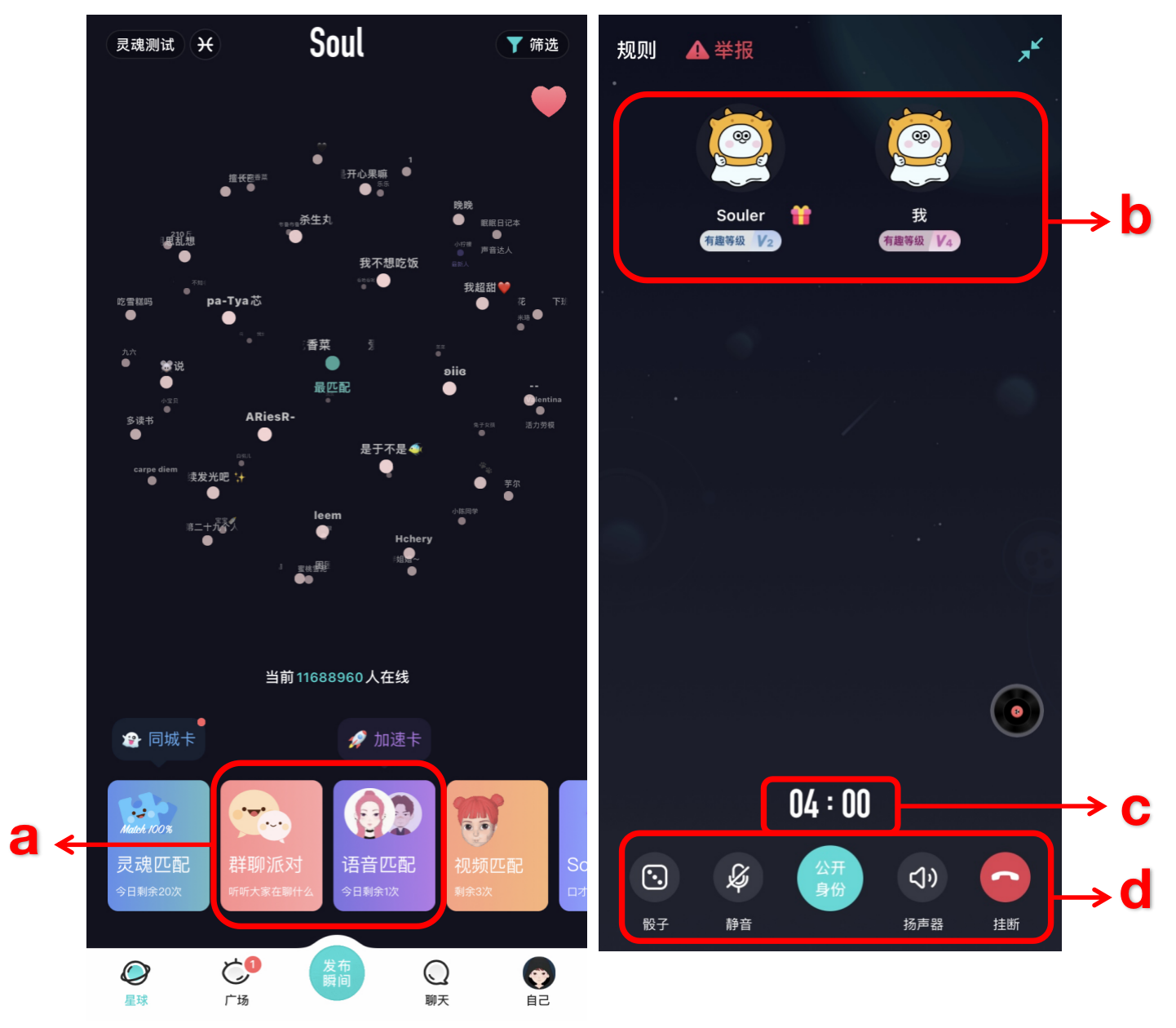}
    \caption{Left: The main user interface of Soul, which contains: (a) voice chat room and one-on-one voice matching. Right: The interface of voice matching, which contains: (b) Display of two users in an undisclosed state; (c) Indication of the remaining time for the chat; the chat will automatically end in four minutes if users choose not to proceed; (d) Various function buttons, including a dice for entertainment, the option to mute oneself, reveal one's profile, speaker controls, and the option to hang up the call.}
    \label{fig:fig1}
    \Description{
    On the left, it shows the main user interface of Soul, containing several buttons that navigate to voice chat rooms and one-on-one voice matching sessions.
    On the right, it shows the user interface of a voice-matching session, which contains the profile images of the two users being matched, an indicator of the remaining time for the matching session, and various control buttons that can achieve different functionalities, such as dice for entertainment, mute oneself, reveal one's profile, speaker controls, and the option to end the session.
}
\end{figure}

In our study, we observed that Soul's emphasis on voice-based matchmaking has the potential to influence user behavior in several ways. Firstly, it directs user attention away from a primary focus on physical appearances, creating an environment that prioritizes communication. This shift appears to provide a more comfortable space for individuals who may have concerns about their physical appearance, potentially encouraging deeper conversations related to personalities, life experiences, and shared interests~\cite{sevi2018exploring}. Additionally, voice communication allows users to convey responses and emotions through intonation and speaking style, which can enrich the quality of exchanges and add depth to conversations.
Furthermore, the use of voice communication on Soul fosters a sense of naturalness and authenticity in interactions. With less reliance on visual elements like photos and text, users may perceive these interactions as more genuine and personal, potentially leading to more positive online dating experiences.

Secondly, the inclusion of voice chat rooms within Soul plays a pivotal role in the quest for potential relationships. These rooms typically feature a host and multiple users participating in synchronous group voice calls. The presence of voice chat rooms diversifies the online dating landscape, offering users additional avenues to discover potential partners. These rooms come in various types, such as speed dating rooms, topic-based rooms, and general rooms, allowing users to join based on their interests and preferences. This further enriches the online dating experience on Soul. Inspired by Soul's popularity, we ask the following research questions:
\begin{itemize}
    \item \textbf{RQ1}: What are the underlying motivations that drive users to choose voice-based online dating platforms?
    \item \textbf{RQ2}: What new affordances do voice-based interactions provide for self-presentation \add{or} self-disclosure \add{separately} in online dating?
    \item \textbf{RQ3}: \add{ How voice-based dating may benefit users under the cultural context of China?}
\end{itemize}

\add{Our research questions are relevant to the \textit{hyperpersonal model} as they focus on the specific aspects of voice communication in online environments and its impact on relationship formation and social behavior. Our findings provide insights into the dynamics of voice-based dating within specific cultural practices. These include: 1) the influence of voice as a moderator in shaping perceptions and managing impressions between the sender and receiver in communication; 2) the advantages that synchronous voice-based and community-based dating platforms provide to users within the Chinese cultural context. Our research contributes to the understanding of online dating communities that utilize voice as the primary modality, specifically within the unique cultural context of China, an East Asian country. }

\delete{Our research contributes a nuanced understanding of real-time voice-based dating communities particularly within a non-Western context. 
This investigation can be framed and analyzed using the Hyperpersonal model, especially within the realm of MDAs. Additionally, we delve into how voice-based interactions can simultaneously safeguard users' privacy and bolster the credibility of the information exchanged.
Furthermore, our findings offer valuable design insights for HCI researchers. These insights can inform the design of user-friendly interfaces and features that leverage the advantages of voice-based social interactions, ultimately enhancing support for match-making and other social activities within MDAs.}



%% file: 01-related-work.tex
\section{Related Work}
Our review centered on various aspects, including self-presentation, the design of online dating features, and the dynamics of dating communities. Additionally, we extended our investigation to encompass online dating cultures in non-Western countries, \add{especially China}, seeking a more profound comprehension of user behavior within the context of Soul.

\subsection{Self-Presentation and Self-Disclosure in Online Dating Sites} 
In the early stages of self-presentation on online dating platforms, individuals aim to establish a favorable initial impression among other users. Impression management involves two key processes: first, the motivation to create a particular impression, and secondly, the actual construction of that impression~\cite{ward2017you}.
There are various strategies to construct an ideal impression, such as crafting an appealing personal profile and engaging with others while adopting a curated persona~\cite{ward2017you, faqs.org}.

One approach to understanding deception in online dating is through signaling theory, which examines the connection between signals and the characteristics users wish to convey. Users express traits that may not be immediately evident and attempt to persuade others to perceive the credibility of these signals~\cite{donath2007social, spence2002signaling, Ellison2017managing}. Previous research suggests that individuals who emit specific signals on online dating sites tend to be more appealing to other users~\cite{datingrose}.
Personal profiles serve as a consistent identity on the platform, aimed at attracting the attention of others~\cite{toma2006examination}. On many online community platforms, creating a personal profile is essential for signaling~\cite{uski2016social}.
To enhance their self-presentation and attract potential matches, users on online dating sites often provide more misleading information compared to other social media platforms~\cite{Masden2015onlinedating,hancock2009putting}. 

In traditional online dating apps, text-based communication is the primary mode through which potential partners interact. In essence, communication methods like text play a crucial role in facilitating interaction and self-presentation among potential partners~\cite{finkel2012online}.
However, voice can also be a significant factor in socializing with strangers, extending beyond dating applications. For instance, in emerging paid gaming teammate communities, multiplayer gamers often select their teammates based on their preferred voice and are willing to pay to play with them~\cite{ShenPaidTeammate}.

The hyperpersonal model ~\cite{walther1996computer} suggests that Computer-Mediated Communication (CMC) has the potential to facilitate more favorable social interactions compared to face-to-face encounters. This model has been applied within the context of online dating to elucidate why interactions in online settings may surpass those occurring in face-to-face scenarios ~\cite{anderson2006predictors, hancock2009putting,ramirez2015online}. The underlying reason, according to the model, is that message senders can effectively utilize the concept of self-presentation.
Furthermore, various CMC technologies, such as instant messaging and online forums, can influence both the type and quantity of information individuals choose to share with others. These choices can significantly impact how people form impressions of one another~\cite{hancock2001impression, walther1996computer}.

The hyperpersonal model comprises four key components: receiver, sender, channel, and feedback, each contributing to enhanced social interaction. \textit{Receiver: }In CMC, receivers interpret messages without the influence of physical appearance and vocal cues, which are typically central to forming impressions. Instead, they focus more on the content and style of the message, often leading to an extrapolation of meaning. They also tend to over-interpret relational messages, attributing more personal interest to online disclosures than in offline settings \cite{jiang2011disclosure}. 
\textit{Sender: }Senders in CMC engage in selective self-presentation. They have the advantage of carefully choosing message elements that best represent the image they wish to convey, a process more controlled than spontaneous speech. This also includes the ability to hide any undesirable physical traits or behaviors.
\textit{Channel: }The communication medium itself allows for further optimization of messages. 
\textit{Feedback: } Idealized perceptions are formed and responded to with self-enhancing messages, which are further optimized through the advantages offered by the communication channel. This process leads to communicators being perceived as, and eventually becoming, more socially desirable.

The hyperpersonal model was applied to the context of online dating to elucidate why users often report greater satisfaction in the online dating environment compared to offline dating. This increased satisfaction can be attributed to the improvement in communication and the reduced reliance on physical cues to establish connections~\cite{anderson2006predictors}.
Online dating platforms offer individuals the opportunity to present an idealized version of themselves~\cite{hancock2009putting,ramirez2015online}. 
Traditional online dating sites often center user profiles around physical appearance, while platforms like Tinder employ the ``swipe right vs. left'' feature, which enables users to form initial impressions of others based on their visual attributes before initiating further communication. This visual-first approach in Tinder exemplifies how online dating profiles are designed in many cases.
Prior research has revealed that the inclusion of photographs in online dating profiles can reduce the idealized impressions that are typically generated by text-only information. This occurs because photographs introduce additional cues related to reality into the communication~\cite{walther2001picture}. On the other hand, text-only information is often perceived as more trustworthy in terms of conveying an ideal impression~\cite{toma2010perceptions}.
\add{Our study contributes nuanced insights to the Hyperpersonal Model, particularly regarding how users engage in selective self-presentation and the formation of idealized impressions in voice-based dating contexts. }


\subsection{Voice-based Social Interaction}

Similar to visual information, vocal cues can convey characteristics about an individual. Voice is interpretable in terms of personality~\cite{yamada2001impression}, and judgments about others' character can be based on their vocal traits~\cite{kramer1970voice}. 
Zhang et.al. conducted interviews with 15 active social media users to explore their preferences regarding the use of synthesized voices to represent their profiles. Their findings reveal that users desire control over how their voice delivers content, including factors like personality and emotion, as these prosodic variations can significantly influence their online personas and impression management \cite{zhang2021social}.
\add{
Research on audio media space is limited.  
Jung et al. conducted interviews with Clubhouse users, a voice-centered social media platform, to understand their motivations, social networking, and conversational habits. They discovered that voice-based interaction effective in forming social relationships \cite{clubhouse}. Ackerman et al. investigated the suitability of audio for shared media systems, focusing on Thunderwire, a high-quality audio media space. They concluded that audio alone can create a functional media space and a valuable social environment. However, they noted that users needed to adapt to various audio-only and system-specific conditions \cite{audioonly}.}

Both vocal and physical appearances can create favorable impressions. This effect is more pronounced when individuals experience only one channel (either seeing physical appearance or hearing the voice) compared to multiple channels (simultaneously seeing physical appearance and hearing the voice)~\cite{zuckerman1990vocal}.
Compared to text-only communication, which relies on users' imagination, people tend to utilize additional physical factors, such as physical characteristics, when evaluating potential romantic partners~\cite{sprecher2002liking}. Online dating apps that incorporate voice communication provide users with tangible physical attributes like reactions, voice tone, and speech patterns to aid in assessing whether a potential partner is a match. 
Sprecher conducted an experiment comparing text-only and voice-only communication in non-romantic relationships, which revealed that parties engaged in voice communication felt a greater sense of intimacy~\cite{sprecher2014initial}. In essence, any form of communication with an audible component tends to outperform text-only communication~\cite{antheunis2020hyperpersonal}.


\add{Limited research in voice communication provides insight into the hyperpersonal model's intricacies. Duthler et. al. conducted a study comparing email and voicemail requests. They discovered that email requests tend to be more polite than voicemails. This difference arises because email allows senders to use more complex language and additional phrases, enhancing the message's politeness. On the other hand, voicemail's inability to be edited limits the sender's capacity to refine their message. The researchers claimed their findings consistent with the predictions of the hyperpersonal model. \cite{duthler2006politeness}.
Another study focused on the gaming community's different reactions to male and female voices. The research revealed that female voices received three times more negative comments compared to male voices or instances where no voice was used. This suggests that voice communication in CMC can sometimes lead to increased hostility or negativity, especially towards women. This phenomenon, termed ``hypernegative effects'', highlights the impact of voice characteristics in online interactions \cite{kuznekoff2013communication}.}


\subsection{Online Dating in Non-Western Cultures}
While dating apps are widely accepted and popular for seeking relationships in Western countries~\cite{castro2020dating}, they often encounter a less favorable reputation in non-Western societies, where negative perceptions about these platforms are prevalent~\cite{blair2019dating,he2023seeking}. This reputational disparity can lead to heightened privacy concerns and a greater tendency to conceal one's personal identity.
Despite developers presenting these MDAs as tools for meeting new people and forming friendships, they are often viewed by many as primarily designed for casual encounters.
MDA users, particularly women, can encounter challenges related to resisting social stigma and dealing with unwanted sexual advances while using these applications.
This divergence in dating and family cultures leads to distinct attitudes toward online dating sites between Western and Eastern cultures. In Eastern countries, the majority of individuals not only consider romantic compatibility but also take into account family backgrounds when connecting with potential partners~\cite{whyte1997fate}. Central to Chinese family culture is the concept of filial piety, where children are expected to date or marry someone approved by their parents. This can sometimes conflict with the use of MDAs, which may go against traditional family norms~\cite{al2017against}.

In non-Western countries, users of MDAs often grapple with heightened privacy concerns due to societal stigmatization and family cultural pressures~\cite{blair2019dating}. To address these concerns and reduce uncertainty about potential romantic partners, users employ strategies to verify the credibility of their matches~\cite{gibbs2011first}.
However, users of traditional MDAs frequently experience a loss of control over their personal information, inadvertently exposing it to unintended audiences~\cite{barnes2006privacy}. This disclosure of personal information in profiles also increases the risk of identity theft or stalking~\cite{gross2005information,spitzberg2002cyberstalking}. Consequently, many users are hesitant to reveal their MDA usage to individuals within their social networks~\cite{rosen2008impact}.
Moreover, for MDAs that rely on image-based matching, users often find it challenging to maintain complete control over who can access their data, further intensifying privacy concerns.

\add{Lots of eastern countries have a collectivistic culture, with China being an example. The collectivistic culture tend to value family, friends and the group over individual. In one study about the social networking site use, researchers found that personal characteristics were less effective in predicting social networking site (SNS) use in China than in the US. This may because in the collectivistic culture, the importance of the family, friends and one’s groups may be partly responsible for Chinese participants’ lesser use of SNSs \cite{jackson2013cultural}.
The collectivist culture in China has undergone significant changes due to economic and social transformations, impacting dating and relationships among young Chinese adults \cite{gaydating2022wang}. There's been a shift from collectivism toward individualism, leading to a preference for romantic relationships over traditional, practical marital relationships. Such transformations motivates the needs in MDA \cite{blair2016dating}. 
While the mainstream of society has a low acceptance on MDAs, Chinese users have a higher privacy concern in MDA use \cite{solis2019meet}. 
In a study focusing on the motivations and risks for Chinese users of MDAs in meeting strangers, researchers discovered that concerns about self-exposure to friends, professional networks, and the community significantly influence their reluctance to meet people offline for casual sex. This highlights the impact of cultural norms and privacy concerns on dating app usage in China, particularly in relation to personal safety and social reputation.}

\subsection{ Online Dating Community and Matchmaking} 
The matching algorithms in online dating apps can result in users being paired with various types of potential partners, and studies have revealed that users tend to select partners who share similarities with themselves, a phenomenon known as homophily~\cite{fiore2005homophily}. This highlights the significance of designing effective matching mechanisms in MDAs to steer users toward potential partners.
Many MDAs are designed with a focus on 'speed dating,' aiming to expedite the matching process and reduce the time spent on communication~\cite{Insider}. However, this emphasis on speed and efficiency may discourage the development of genuine relationships and potentially foster a concentration on physical appearances.
For traditional MDAs, the discovery mechanisms for users typically prioritize geographical location~\cite{zytko2021computer}. Many MDAs also include profile pages and private messaging interfaces as key features to help users screen potential partners~\cite{zytko2021computer,fiore2004online}. However, the common process of ``discovering profiles, receiving match notifications, and interacting through messaging'' often leads to misunderstandings and discrepancies between the personalities of users and how they present themselves in their profiles. Photos and text-only communication may not be sufficient for users to fully understand each other~\cite{Insider}. 
According to Lee and Bruckman, online dating is closely intertwined with real-life dating activities, often mirroring real-world dating dynamics~\cite{lee2007judging}. In addition to one-on-one interactions, online dating can foster a sense of community where users share emotions and exchange information with each other~\cite{wellman1999net,Masden2015onlinedating}.

Traditional online dating apps typically rely on asynchronous and visual communication~\cite{hsiao2012synchronous,gewirtz2018forever}. Asynchronous communication implies that conversations between users may be delayed over time, which often happens in text-based communication. This delay can occur if one party does not respond immediately to a message or if one party is offline when a message is received. Visual communication refers to the fact that traditional online dating users have an initial visual impression of each other before they engage in communication. Successful matches are often based on photos, allowing both parties to form an initial impression of each other's appearance before initiating formal communication.

Soul distinguishes itself from traditional apps by not relying on asynchronous and visual communication. In Soul's voice-matching process, users remain anonymous and do not have access to each other's profiles, eliminating the influence of visual impressions. Communication on Soul takes place through voice, facilitating real-time interaction without the delays associated with messaging. Soul also introduces a unique feature, the voice chat room, where members can engage in group-based voice communication.

%% file: 02-method.tex
\section{Method}

\add{To explore user behavior on Soul, we utilized a semi-structured interview approach with 18 participants. Our methodology focused on conducting in-depth discussions to gain insights into their experiences. Following the interviews, we employed thematic analysis to methodically examine and interpret the data collected. This approach allowed us to identify and understand key themes and patterns in user behavior within the context of the Soul app.}


\subsection{Participants}
For our research, we enlisted the participation of 18 committed users of the Soul app. 
We recruited participants by initiating conversations in topicless chat rooms, entering as regular users to mingle with others. After exchanging greetings and introducing ourselves, we mentioned our ongoing study on the use of Soul and asked if anyone was interested in participating. For those who expressed interest, we arranged to connect with them via WeChat.
To ensure that our sample was representative, we specifically chose individuals who had engaged in several voice-matching experiences and had dedicated more than 5 hours to participating in voice chat rooms on the app. The demographic breakdown of our participants is detailed in \autoref{tab:my-table}.


\begin{table*}
\caption{Summary of Interviewees}
\Description{The table provides a snapshot of 18 individuals, identified by unique IDs ranging from P1 to P18, and offers a glimpse into their demographic and professional backgrounds. The gender distribution is balanced, with a mix of male (M) and female (F) participants. The ages of these individuals are predominantly in the early to mid-20s, specifically between 20 and 26 years old, reflecting a young adult demographic. Occupation-wise, the majority are students, underscoring a strong academic inclination within this group. Additionally, there are a few individuals engaged in professional roles, including intern, teaching, design, and an office clerk position.}
\label{tab:my-table}
\begin{tabular}{cccc||cccc}
\toprule
ID & Gender & Age & Occupation & ID  & Gender & Age & Occupation \\ \midrule
P1 & M      & 24  & Student    & P10 & F      & 21  & Student    \\ 
P2 & M      & 23  & Intern     & P11 & F      & 22  & Unemployed \\ 
P3 & F      & 22  & Student    & P12 & F      & 22  & Student    \\ 
P4 & M      & 24  & Student    & P13 & F      & 23  & Designer   \\ 
P5 & F      & 21  & Student    & P14 & M      & 24  & Student    \\ 
P6 & M      & 24  & Intern     & P15 & F      & 23  & Teacher    \\ 
P7 & F      & 24  & Student    & P16 & M      & 24  & Office Clerk         \\ 
P8 & F      & 24  & Student    & P17 & M      & 20  & Student    \\ 
P9 & F      & 26  & Teacher    & P18 & M      & 24  & Student    \\ \bottomrule
\end{tabular}
\end{table*}

The active user base of Soul is typically comprised of individuals under 35 years old~\cite{jianshu}. To ensure that our participants fell within this age range, our primary focus was on individuals aged between 18 and 35 during the recruitment process. Our final sample consisted of 18 participants, with ages ranging from 21 to 26, including 10 female participants and 8 male participants. \add{Out of 18 participants, 1 had experience hosting a voice chat room.} All participants were informed that the interview process would be conducted anonymously. When selecting participants, we adhered to the following criteria: first, all participants had to be dedicated Soul users with over three months of experience, and second, they needed to have experiences with traditional dating apps as well.

\add{\subsection{Procedure}
The interviews were meticulously conducted using Zoom, supplemented by WeChat for some participants, to provide additional insights during the analysis and coding stages. At the beginning of the interview, participants were thoroughly briefed through the reading of a consent form. This form detailed critical aspects such as the nature of the questions to be asked, the methodologies for data handling and storage, and the ethical considerations including participant rights. To proceed with the interview, verbal consent from each participant was mandatory. }

\add{The duration of each interview was approximately one hour. Interview questions were structured in four key areas:
i) motivations behind the participants' use of the Soul application;
ii) experiences and feedback regarding the app's voice matching functionality;
iii) nature and quality of interactions within Soul's group chat rooms;
and iv) comparative evaluations of Soul against other MDAs.
}

\subsection{Data Collection and Analysis}
\add{The interviews for our study were conducted remotely via Zoom and were recorded in video format, generating approximately 20 hours of recorded material. Each interview session was captured as a video recording and later transcribed verbatim for analysis. One of the interviewers actively used the Soul app, providing an insider perspective during the research process.}

In the analysis phase, three out of the four researchers employed thematic analysis ~\cite{corbin2014basics} on the transcribed text. The team's roles were well-defined: one member was tasked with transcribing the video recordings into text format, while the other two researchers focused on coding and reviewing the transcribed data. Our thematic analysis combined both inductive and deductive coding strategies.

The analytical process encompassed several critical steps. Initially, each reviewer independently generated codes from their assigned section of the transcript, resulting in an average of about 50 unique codes per transcript. These initial codes included terms such as ``fear of being recognized by surroundings'', ``imagination (identity)'', ``imagination (personality)'', and ``stereotypes in voice/speaking style''. 
Subsequently, we identified the relationships between the codes and grouped these codes deductively into broader categories, such as ``privacy concerns'', ``receiver envisioning sender'' and ``perceptions of voice''. We considered previous literature and theories, such as Hyperpersonal Model when developing the coding scheme. 
The final step of our analysis involved identifying relationships between these categories to develop a comprehensive set of themes. This process led to the formation of 30 distinct categories, which included both inductive insights and deductive groupings.

\delete{This process involved assigning specific codes to various segments of the text, subsequently organizing and categorizing these codes into distinct themes, following Corbin and Strauss's guidelines~\cite{corbin2014basics}. 
The coding process was a collaborative effort among a team of three researchers. To ensure precision and reliability, all codes underwent thorough examination and analysis through multiple iterative rounds involving the three researchers.}

\subsection{\add{Ethical Considerations and Positionality Statement}}
\add{This research project, including all related materials (e.g., recruitment plan, consent form, and interview scripts), was reviewed and approved by the Ethical Review Board prior to the commencement of the study to ensure compliance with ethical guidelines and research integrity. Before participating in the interview, each participant was provided with a detailed consent form to read and comprehend, ensuring informed consent.}

\add{The research team acknowledges the potential influence of our personal genders, cultural backgrounds, and social identities on the study, recognizing the possibility of introducing biases. Our team comprises four researchers: three identify as men and one as a woman. Ethnically, we are all of Asian descent. In terms of our relationship with the platform ``Soul'', our experiences vary: one team member is a dedicated Soul user, one has used Soul once, and two had not used Soul before this project. This diversity in our experiences with the platform is acknowledged as a factor in our research approach.}

\section{The Mechanics of Soul}

Soul provides two avenues for users to discover potential partners: voice matching and group chat rooms (\autoref{fig:fig1}). Voice matching entails users being matched with potential partners based on their personal information and preferences, followed by engaging in voice communication to establish a connection. Conversely, group chat rooms enable users to partake in group discussions on diverse topics, potentially leading to encounters with potential partners through these conversations. 


Similar to traditional dating platforms, Soul requests users to furnish fundamental details about themselves during the registration process. This information typically includes gender, interests, and personality traits. These details serve as the basis for Soul's matchmaking services. When a one-to-one pairing is established, both users' profiles remain concealed from each other during the initial five minutes of the voice chat. During this time, their sole means of evaluation is through their voices. If both participants express mutual interest in extending their interaction, they have the option to unveil their profiles to each other, granting them unrestricted access to voice communication.

Soul offers users a choice of three distinct types of voice chat rooms: speed-dating rooms, topic chat rooms, and topicless rooms.
\begin{itemize}
    \item Speed dating rooms: The purpose of these rooms is to assist users with particular preferences in discovering potential matches. Typically, these rooms are overseen by a \textit{host}, a dedicated user responsible for creating the group chat, moderating the conversation, and steering the discussion toward a specific topic. The host listens to users' preferences and aids in identifying suitable matches based on those criteria.
    \item Topic chat rooms: Revolve around specific subjects, like hobbies or books, and participants in these rooms engage in discussions related to that particular topic.
    \item Casual chat rooms: Open-ended, allowing users to engage in discussions about a wide range of topics without any specific theme or constraint.
\end{itemize}
The voice chat rooms typically consist of four main components: the topic, the host(s), the speaker(s), and the audience(s):
\begin{itemize}
    \item Topic: Only the topic chat rooms and speed dating rooms has a topic. A valid topic can be a hobby or a shared experience.
    \item Speaker(s): The participant in the voice room who is allowed to open their microphone and speak is typically the person looking for potential matches. 
    \item Host(s): A experienced user who manages the voice-chat room and help participants get matched. The host decides the topic of the room and who can be the speaker.
    \item Audience(s): Everyone else in the voice-chat room. 
\end{itemize}
These elements interact to shape the dynamics of the group chat room within the voice-based online dating app Soul, as illustrated in \autoref{fig:fig3}. 
\delete{In the upcoming sections, we will delve deeper into the dynamics of these various voice chat rooms on Soul.
For further insights into voice chat rooms, please refer to section 4.3, where we delve into additional findings.}

  \add{In our interviews, participants discussed using self-created voice tags on the Soul platform to aid in finding romantic partners. These tags, developed by the users themselves rather than Soul's product team, are not used for automatic matching or filtering. Instead, users independently create and use these tags. To find someone with a desired voice type on Soul, users join voice chat rooms to listen to hosts or participants. Hosts may highlight a user's voice type to attract attention. Additionally, speed-dating rooms on the platform often welcome users with specific voice types.}

 \add{
 The Soul community values the use of imagination in voice interactions, enhancing conversational engagement. With limited visual information, users rely on voice to understand each other's identities. To help form mental images, descriptive labels for voice timbres are used, such as ``middle-aged'' or ``youthful'' for male voices, and ``authoritative'' or ``girlish'' for female voices. These categorizations aid in comprehension and connection, shaping user perceptions and interactions on the platform.
}

%% file: 03-findings.tex
\section{Findings}
In the upcoming sections, we will delve into our findings and examine the factors motivating users to embrace voice-based dating platforms. We will also explore the distinctive dynamics of voice-based community interactions, and investigate the role of perception and regulation within these voice interactions.

\subsection{Motivations Behind the Use of Soul}
\subsubsection{Multi-modal Information Needs}
Similar to users of other dating apps, the primary motivation for Soul users is to satisfy their social interaction needs. Voice communication offers users the opportunity to glean more information about their potential matches than they might from a meticulously crafted profile page. Since users are unable to access each other's profiles until a few minutes into the voice chat, they depend on their voices to create an initial impression. Voice timbre, speaking style, and other attributes conveyed through voice communication collectively contribute to this initial impression. As articulated by P3, 
\begin{quote}
\textit{``It's easier for me to evaluate one person through voice chat. I used to meet someone on Tinder who identify himself as a highly educated person. We have good conversations by texting to each other. Once we had a voice chat, I noticed that he wasn't the person he pretended to be immediately because of the way he spoke. Chatting through voice helps me better discover what the person really be like.''} (P3)
\end{quote}

Some participants expressed that they would invest more effort in sustaining a conversation if they found the other user's voice pleasant. For instance, P7 recounted her experiences using other dating apps in contrast to Soul, stating, 
\begin{quote}
  \textit{``I used to use Tinder. I would not spend the effort to keep the conversation via text messages and will not feel bad about pausing it. On Soul, I will take the initiative to continue the conversation if I like the person's voice and talking style.''} (P7)  
\end{quote}


\subsubsection{Voice Communication Enhances Privacy}
In our study, six female participants underscored the negative reputation associated with traditional dating apps, noting that a significant number of individuals view them as disreputable. They articulated the belief that Soul offers a safeguard for their privacy by initially concealing their physical appearance.
P5 desired to present her authentic self on dating sites while avoiding recognition by individuals who knew her personally. However, a text-based dating platform that required personal images in profiles did not align with her requirements,
\begin{quote}
    \textit{``If people around me know that I am using this kind of app, it may damage my reputation. If I put my image on this kind of app, there's a chance that I could be recommended to people who know me...You have to upload a personal image in some apps. I wouldn't upload a fake selfie since I don't want to engage with someone who uses a fake selfie either.''} (P5)
\end{quote}
Voice-based dating apps, in comparison to text-based dating apps lacking profile images, offer a higher degree of authenticity while simultaneously providing enhanced privacy protection. P2 and P4 expressed the similar opinion. As mentioned by P4, 
\begin{quote}
    \textit{``I want to reveal part of my real self without revealing my identity to people who may know me personally. Voice-based dating platforms can strike this balance.''}
\end{quote}

\subsubsection{Emotional Enrichment Through Voice Communication Heightens Intimacy}
Some participants suggest that they could receive and express a richer emotion signals via voice communication.
Some participants felt it was easier to express their attitudes to potential partners via voice compared to text, \textit{``I don't need to meticulously craft my messages to avoid misinterpretation,''} as P19 emphasized, \textit{``I simply express my feelings naturally in the conversation. It make things easier.''}
Some users may get more emotional and empathetic to others by having voice calls instead of text messages on Soul. The emotions in one's voices can trigger one's desire to continue the conversation,
\begin{quote}
    \textit{``Once I match with someone on the app who is using it due to loneliness and a desire to find someone to talk to. I would immediately pause the conversation if it's through text because I consider it a waste of time. However, when I hear their voice, I can sense their loneliness, and it reminds me of how I felt when I had just graduated from college.''} (P15)
\end{quote}

\subsection{Perceptions and Moderation in Voice Interactions}
\add{In our study, we found that during voice exchanges, users often prioritize the \textit{timbre} and \textit{style} of the other person's voice over the actual content of the conversation initially. This tendency allows them to form an impression of the person's personality and attitude based solely on their voice. We adopted the framework of the Hyperpersonal Model, defining speakers as receivers and listeners as senders. Our research highlights how both senders and receivers perceive each other through voice, or use their own voice for self-presentation. This is based on the concept of forming a \textit{mental image} of a person through their voice. Additionally, we explored the role of group voice chat room hosts in facilitating and moderating these voice interactions.}




\subsubsection{\add{Receiver: }Envisioning the Individual via Their Voice}

\add{The user assumes the role of an information receiver when they are in the position of a listener.} As voice is the central mode of interaction on Soul, users heavily depend on it to shape their perceptions and assumptions about each other's identities. The voice timbre, in particular, serves as a potent cue that enables users to visualize the appearance and identities of others based on voices and real-life experiences. The cue can be timbre, accent, speech mannerisms, and tone. For instance, if a male user's voice has a ``teen boy's'' timbre, \delete{fellow users} P3 would envision him as a college student. P4 mentioned that he tends to imagine a girl who is taller than him whenever he hears a female speaker with an accent from the northern regions of China, ``they usually glow taller than use from south'', he said.

\add{When users hear a voice type they like, they tend to idealize the speaker. P16 suggested that based on personal experience, it's easy to imagine a specific type of ideal partner just from their voice. \textit{``I've realized I particularly like a certain type of girl. In the past, I paid extra attention to girls of this type in real life and noticed a similarity in their voices,''} he said. When asked about the mental image he forms, P16 became a bit shy,
\begin{quote}
 \textit{``You know, I like girls who speak slowly and gently, sometimes a bit abstractly... Then, I picture them as well-educated ladies, likely majoring in arts or social sciences... I often imagine them in the same way.''}    
\end{quote} 
P14 shared his experience about how people find him attractive because of his voice, \textit{``Many girls I met on Soul told me that because of my soft voice, they would imagine that I am a handsome gentleman, which makes them more willing to spend time chatting with me. ''} Similarly, P13 noted that the lack of visible profile in one-to-one pairings fuels her imagination, \textit{``Everything I know about him comes from his voice, which lets me imagine all the positive possibilities about him. Such imagination motivates me to continue the conversation and learn more about him.''} She also compared voice-based dating to text-based dating apps and highlighted Soul's advantages in online dating, \textit{``It’s more intuitive, and the information is richer, allowing me to build up a mental image of this person in just a minute.''}} 

\add{When asked how their mental images would evolve through further communication, and whether an idealized image could negatively impact their perception if it contradicted the sender's real identity or personality, both P13 and P16 said the biggest impact of the voice type was encouraging them to invest more time and effort in getting to know the sender. P16 mentioned he is open to modifying his initial perceptions based on ongoing conversations, \textit{``If her voice initially leaves a good impression, I find it worthwhile to invest more time and effort to understand her better and include more details to imagine her personality.''} On the other hand, P13 noted that she tends to let go of any idealized mental images formed based on the sender's voice once deeper conversations begin, \textit{``His voice is what motivates me to engage in further conversations with him. But I believe the best way to truly know someone is to avoid making assumptions from the start. ''}}

These voice types serve as a convenient tool for users to convey their preferences for specific voice timbres in potential partners. In scenarios like the speed-dating room, a user can simply reference a desired voice type, allowing other users to quickly understand the type of voice timbre the user seeks in a potential partner. This streamlined communication aids users in effectively connecting with each other on the app. 

\add{Additionally, beyond forming impressions of each other's identities and personalities through voice, receivers can also discern the sender's attitude from how they speak. P3 shared her experience of evaluating male users' attitudes in the speed dating room, \textit{``I can sense their emotions from the way they talk. It's hard to describe, but I can feel it. Some talk arrogantly, making me feel as if they're doing me a favor by speaking to me.''} P7 commented on her ability to gauge attitudes in topic chat rooms, \textit{``If I sense that the discussion is turning into a debate, I'll leave the room immediately. People's tone changes when the conversation is no longer peaceful.'' }}
\delete{personalities. This imaginative process often incorporates an element of idealization, as users who find satisfaction in the other person's voice may be inclined to imagine that the individual possesses positive characteristics. As mentioned by P14, ``Many girls I met on Soul told me that because of my soft voice, they would imagine that I am a handsome gentleman, which makes them more willing to spend time chatting with me.''}

{\subsubsection{\add{Sender: Managing Self-Presentation via Voice}}
Some users may deliberately modify their voices as a strategic attempt to appear more appealing to potential partners. This alteration might include using vocal techniques to adjust the timbre or pitch of their voices or adopting specific speaking styles they believe will enhance their attractiveness to others. As mentioned by P2.\textit{``I will lower my voice and speak more slowly when talking to people on the soul. I want to leave the impression that I am a gentle person with a delicate voice. ''} 
\add{P7 shared that she intentionally makes her voice sound ``cute'' by speaking in a higher pitch and softer tone, and by giggling when she laughs. She noted that the impression
she creates in voice chats contrasts with the first impression she typically gives in person,
\begin{quote}
\textit{ ``I'm taller and larger than most girls around me, and many people I've met in life don't see me as someone they need to protect. At some point, I started acting more like a tomboy and felt embarrassed about any feminine behaviors.''} (P7) 
\end{quote}
P7 also discussed how voice-based dating platforms allow her to present herself in the way she prefers.\textit{ ``I want to be treated like one of the delicate and soft girls, so I speak with a pitched voice. I can talk the way 'soft girls' talk.''} Most importantly, she feels more comfortable presenting herself on voice-based dating apps without showing visual information. \textit{``I would feel embarrassed and ashamed if people saw my photo, because my appearance doesn't fit the typical cute and feminine girl, but that's how I want to be treated.''} }

\add{Some users experience a lower sense of self-deception when they modify their voice and speaking style to present themselves, compared to editing their selfies for dating profiles on other platforms.\textit{ ``I can choose to speak in that manner anytime, but I can't always look like my edited selfie. Sometimes I feel like I'm lying to myself about how I really look,''} P7 explained. P8 shares a similar viewpoint, \textit{``I've used Tinder before and posted an edited selfie. I didn't feel too bad about it because it seemed like everyone was doing it. However, it always feels better when people are attracted to the real you.''}}

However, there also exist users who use vocal changer software to decorate their voice, \textit{``I used to talk with someone whose voice sounds unnatural in a voice chat room,''} said by P17, \textit{``I think she was using a vocal changer software, but it only makes her voice sound weird. It also can be a male who uses software to pretend to be female and make fun. Anyway, I don't like it.''}


\subsubsection{Host—The ``catalyst'' for group chat room}
The host plays a pivotal role in group chat rooms, facilitating interaction, guiding conversations, and helping users find compatible partners. They also regulate who can speak to maintain focus and efficiency. In topic chat rooms, hosts initiate conversations and set topics to foster connections. They moderate discussions to ensure relevance and respect within the community.

The first role of the host is to facilitate efficient dating in speed dating rooms, which typically have a large number of users looking for quick connections. The host acts as an organizer, screening suitable users and promoting efficient dating within the room. \textit{``The host needs to drive the vibe and organize discourse in all aspects. ''}(P9)

\add{The second responsibility of the host involves cultivating a sense of community and encouraging the development of friendships in various group chat rooms. }
\delete{Secondly, the host can help foster a sense of community and friendship.}
Users often frequent the same group chat rooms, sharing their experiences and getting to know each other over time. Typically, when a host creates a room, a consistent group of members joins, and they gradually become familiar with each other, forming a social circle. The host regularly sets up group chat rooms, which serve as the foundation for building long-term friendships. As mentioned by P9,
\begin{quote}
    \textit{``There is a user who is the owner of the group chat room. By chance, I entered his group chat room because his voice timbre is very pleasant. He opened the room every day. At that time, the other two girls included me entered his group chat room on time every day...\add{and we became friends.}''} (P9)
\end{quote}

In topic chat rooms, the host also facilitates connections between users. These chat rooms typically have around 4-5 participants, depending on the host's settings when creating the room. Unlike speed dating rooms, chat rooms have a more relaxed atmosphere, and users may already be engaged in lively discussions. When a new user enters a topic chat room, the host often invites them to use their microphone and encourages them to join the ongoing conversation. This allows new members to participate in the topic discussion, providing an opportunity to connect with potential partners or friends who share the same interests.
\begin{quote}
    \textit{``Basically, in order to seek politeness, the first reaction of everyone is not to interrupt rashly, but to listen for a while and then interrupt... If the host is in, he may invite users to speak if he finds a new partner coming in. Then at this time, users are not good to refuse, so that its retention rate will be higher. So actually the host I feel quite necessary.''} (P1)
\end{quote}

Furthermore, most hosts in Soul group chat rooms are expected to have a pleasant voice, as they play an active role in moderating and managing the room. A host with an appealing voice can attract more users to stay in the room and increases the likelihood of user engagement and participation.
\textit{``Many people will stay in the room because of the voice timbre of the host. The good voice timbre of the host can attract more people to enter the room.''} (P12)

\subsubsection{The Experience of Voice Interaction} 
Voice communication allows users to perceive the natural responses of the other party, enabling them to decide whether to continue the conversation based on these signals. Users can gauge the other person's interest and engagement in the conversation, which helps them determine whether to pursue further interactions. As participant P18 mentioned, \textit{``Using voice to communicate makes me feel real because we can directly hear the other person's words, tone, and intonation.''} When one person shows interest in the other, their tone of voice may convey a positive attitude. These emotions can be detected through the voice, with changes in pitch, volume, and other vocal cues indicating the speaker's emotional state.

In text chat, users have the option to delay or pretend not to have seen messages they do not want to respond to immediately. However, there is no such delay in Soul's voice communication, and users must respond promptly. This immediacy makes it challenging to maintain long-term impersonation. Additionally, the synchronous nature of voice communication enables users to quickly reject someone who makes excessive demands. P15 mentioned that she would hang up when encountering users she did not like.
\textit{``If someone keeps mentioning that he wants to see my photos during the voice dating process, I will end the chat and leave immediately.''} (P15)

\subsection{Various Voice-Based Community Interactions and their Dynamics}
In addition to one-on-one voice chat, the voice-based dating platform we examined also provides group-based voice chat environments. This allows us to explore how users engage within a \add{well-established,}\delete{ matured} voice-based dating community and gain insights into their perceptions of such interactions. There are three types of group chat rooms available on Soul: speed-dating rooms, topic chat rooms, and casual chat rooms.

\begin{figure}
    \includegraphics[width=8cm]{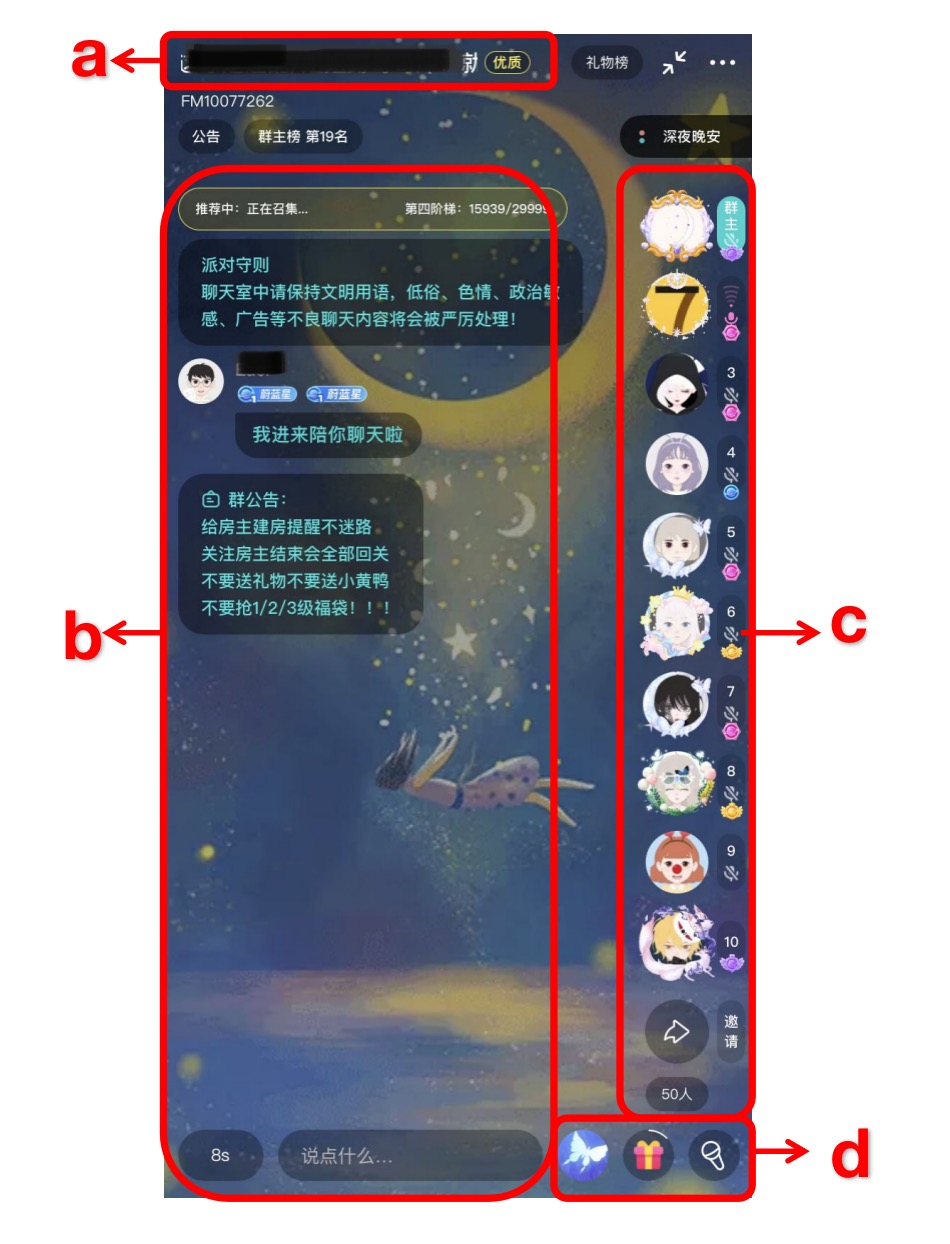}
    \caption{The interface of the Group chat room contains: (a) the topic of the Group chat room, if there is one; (b) the text chat area; (c) the top icon is the host user, and the rest users are divided into users with and without the mic. The users must apply to the host to open the mic and talk in the room. The discrimination method is to see whether there is a microphone sign on the right, and the number of people in the room is displayed at the bottom. (d) function buttons, which enable users to trigger an animation (for entertainment), send virtual gifts, and apply for permission to open up the microphone.}
    \label{fig:fig3}
    \Description{The screenshot of the interface of the group chat room. The top area of the interface contains information about the topic of the group chat room if the users set it. The left side is an area where users can chat via text messages. On the right, it shows a list of users who are currently in the chatroom, showing their profile images and badges. On the bottom right corner, there are three buttons by which users can trigger animations for entertainment, send virtual gifts to other users, and open the microphone to talk.}
\end{figure}

\subsubsection{Matchmaking with Complex Requirements with the Help of a Host \add{in Speed-Dating Room}}
In Soul's speed-dating room, the goal is to help users with specific and diverse preferences quickly discover potential matches, all under the guidance of a host. Here's how the process works: Female users take turns to be the speaker by activating their microphones and describing the characteristics of their ideal partners. Male users who believe they meet these requirements can request to speak with the female user. The female user then chooses one male user, and they leave the room to initiate a private conversation. Throughout this process, the host moderates the conversation and determines who can use their microphones to address the group. Typically, the entire matching process is completed in less than two minutes.

Utilizing a speed-dating room can be more efficient in comparison to one-to-one matching on Soul. The advantage lies in the user's ability to engage with multiple potential partners simultaneously and assess them concurrently with minimal effort, e.g.,
\begin{quote}
\textit{``You can meet multiple people and talk to them simultaneously. Since it is a group conversation with the help of the host, you can get to know them quickly without spending effort on keeping the conversation going''}. (P7) 
\end{quote}

In the matchmaking process in speed-dating room, the voice is a important factor. Users often appraise each other by their voice timbre and talking style. Since the time is limited,  
Therefore, voice timbre is still an important factor in speed dating room. Because the speed of making friends in a speed dating room is very short, and the speaking style and personality are difficult to be shown in a very short time, then users can only produce the first impression through the voice timbre, and often good voice timbre will gain more attention. As mentioned by P9, \textit{``All the girls or boys in the speed-dating room will have requirements. Generally, ninety percent of them hope that the other party has a pleasant voice timbre.''}


\subsubsection{Finding Romance through In-Depth Conversations in \add{Topic Chat Room}}
The topic chat room on Soul is distinguished by its specific and in-depth themes compared to a casual chat room. Conversations within a topic chat room revolve around a particular subject, allowing users to establish personal connections with others during the discussion. 
\add{The room's host typically sets the topic, guiding the direction of the dialogue.}
Topics in these chat rooms can vary widely, encompassing shared experiences or discussions about movies, for instance. Among these rooms, the emotional room is the most popular, where users share their past relationships, provide support, and offer advice to one another. According to P11, \textit{``The people who share their past relationship are looking for comfort. Some people want other users to critique their ex-boyfriend or girlfriend together with them. However, the ultimate goal of using the emotional room will always be looking for a new relationship.''}


The voice chat room provides a structure and purpose for the conversation, which may be lost when the conversation moves to a different platform.
It's common for online dating users to connect on third-party social platforms like Instagram or WeChat after getting to know each other on the dating app. However, in our study, we observed that some users prefer to limit their interactions solely to the topic chat rooms on Soul. When conversations lack a specific topic, they can sometimes become stagnant or come to a halt. For instance, P1 shared an experience where members in a topic chat room ceased chatting after attempting to continue their conversation on WeChat, \textit{``After transferring the conversation to the WeChat group, some people started talking less, and eventually nobody talked anymore.''} P12 participant suspected that it may because it's easier to initialize voice chat on Soul,\textit{ ``It's common to have voice calls on a voice-based dating site like Soul. It can feel more intimate and even inappropriate to request a voice call when using other platforms.''} 
\delete{P15 pointed out that after transitioning to a different platform, users lose the flexibility to join and leave rooms as they wish, stating, ``We can enter and exit rooms at any time we want, and we know what topics to discuss. 
If we switch to a different platform, it feels somewhat more restrictive.''
}


\subsubsection{Connecting with Others through \add{Casual Chat Room}}
In casual chat rooms on Soul, the host facilitates the interaction, and participants can join the discussion at any time by activating their microphone. These chat rooms lack a specific topic and allow users to engage in free-form conversations. The atmosphere in casual chat rooms is typically relaxed, and users are often evaluated based on their speaking abilities and sense of humor rather than their voice timbre. It's crucial for users to be mindful of the chat room's ambiance and purpose and adapt their communication style accordingly.
\textit{``In casual chat rooms, I think it is acceptable if the voice is not bad, and interesting words are more attractive. Compared with speed dating rooms, casual chat room people are relatively few. Chat time will be longer.''} (P13)


\subsection{Negative Experiences with Soul}
Soul, like many other online dating apps, also grapples with a common issue—user fatigue from seeking new connections. Several participants revealed that they turned to online dating apps out of boredom, using them to combat loneliness and pass the time. However, this very convenience of always having new people to meet can lead to user fatigue, as individuals may become restless and develop unrealistic expectations that the next person they encounter will be even better. As P6 explained:
\begin{quote}
    \textit{``It's a bit tired online dating because you will meet so many people. You feel like you're always going to meet someone better, and you are always picking and choosing, which can make you tired.''} (P6)
\end{quote}

\subsubsection{Stereotypes on Voice}
Soul, as an online dating app centered around voice communication, places a significant emphasis on the impression created by a user's voice. \add{As we mentioned in Section 5.3,} users form diverse impressions of others based on their voices.
\delete {as discussed in our voice analysis.} 
Individuals with melodious voices often have an advantage in making positive impressions, increasing their chances of meeting potential partners.
However, it's crucial to acknowledge that not everyone possesses a pleasant or refined voice. Soul may not provide an equally favorable experience for users with average or less appealing voices. Some users with non-standard accents also faced discrimination on the platform. Participants raised concerns about accents, generally associating them with lower education levels. As mentioned by P9,
\textit{``Because I was in a speed dating room. I often meet people who require to match with people that speaks standard Mandarin without accent.''}
P5 also holds the same point, \textit{``For example, if the person does not speak standard Mandarin and then speaks with a strange tone, I will not have the desire to make friends with that person person.''}

\delete{The discrimination against non-standard accents is rooted in the rise of short video platforms in China over recent years. These platforms have seen an influx of content that may include crude language and unrefined Mandarin, often presented through humorous or attention-grabbing spoof videos. This trend has led to negative associations between lower education levels and poor Mandarin skills in the minds of many.
However, }
It's important to note that regional differences naturally result in various accents across China. \add{What one person in the north might regard as non-standard Mandarin could be perceived quite differently by someone from the south. }
\delete{What one person may perceive as non-standard Mandarin in the north, someone from the south might consider the same way.}
\add{Therefore, such perceptions of ``standard'' versus ``non-standard'' Mandarin are often subjective and influenced by regional biases. The underlying cause of accent discrimination can be attributed to the varied environments in which people are raised, leading to different linguistic experiences and perceptions \cite{spence2022your}.}

\subsubsection{Conformity Phenomena and Voice}
Some users become overly focused on voice timbre and may attempt to intentionally alter their own voices. Due to the popularity of certain timbres, like the bass voice, individuals might strive to sound more like these favored timbres. However, this deliberate imitation can have unintended consequences. One female participant mentioned encountering male users who intentionally deepened their voices to mimic the bass type. Participants shared that these attempts often came across as insincere and forced, leading to negative perceptions of those individuals.
\begin{quote}
    \textit{``I think some boys speak a little affectation. But I don't know how to put it there are some people who start out just trying to attract people maybe or something. Because some girls, they just hang up when they don't like it. So some boys may talk deeper and more magnetic at the beginning to attract people. When we get to the end of the conversation, he relaxes. The voice may not be as good, or as magnetic.''} (P5)
\end{quote}

\delete{Striving for popularity is a common phenomenon, particularly in online dating apps. 
However, it's essential for each user to embrace their uniqueness, as everyone has their own distinctive qualities. There's no need to conform to the same standards and risk losing one's individuality \cite{frost2008people}.}


%% file: 04-discussion.tex
\section{Discussion}
Our research aimed to investigate the impact of real-time voice-based MDAs in \add{China }\delete{a non-Western context}. We explored the process of forming and nurturing potential romantic relationships through one-to-one voice interactions and group chats.

\subsection{Affordances of Real-Time Voice-based Interactions in Online Dating}
The Hyperpersonal Model proposes that individuals tend to be more satisfied with online dating because they disclose less information about their potential partner, which can lead to idealizing that person.
\add{Our research supports the hyperpersonal model, particularly in the context of voice interactions. In these scenarios, the roles of the receiver and sender can be seen as fluid, with speakers acting as receivers and listeners as senders. }

\subsubsection{\add{The Role of Voice in CMC for Dating and the Influence of the Host.}}
\add{
In previous research examining the online dating community, researchers identified a need for more ``novelty" in online dating \cite{Masden2015onlinedating}. This refers to users of online dating forums expressing a desire for more enjoyable and diverse forms of socialization, as opposed to the uniformity often found in these platforms. We propose that voice-based dating addresses this need for novelty by introducing a different form of communication, thereby diversifying the user experience in the online dating landscape.
Based on the real-time nature of voice CMC, we suggest that voice-based dating may enable users to form a mental image of their potential partner more quickly than text-based dating, due to the rich information conveyed through the voice channel. 
Additionally, receivers who have a mental image of an ideal partner may quickly associate the sender with this ideal when they hear a voice that matches their model. Previous research indicates that voice communication is often perceived as more trustworthy than text because of its richer cues \cite{jiang2013perception, sillence2004integrated}. However, in the context of online dating, senders might intentionally alter their voice and speaking style for self-presentation. 
Therefore, we hypothesize that receivers are more likely to accept the sender’s crafted self-presentation on voice-based dating sites than on text-based dating sites.}

\delete{Unlike many other MDAs that offer built-in voice chat but face user hesitancy for various reasons such as self-consciousness about their voice, privacy concerns, or personal preference for typing, Soul takes a distinct approach. It is purposefully designed as a voice-based MDA, where voice chat is the default mode of communication. This intentional design choice aims to eliminate psychological barriers, making users more comfortable with using voice communication and encouraging them to engage in real-time conversations.}

\add{The presence of voice types on platforms like Soul plays a significant role in fostering the idealization of voices. Our investigation into Soul's platform reveals that the most commonly used labels for male voices are "teen-boy" and "bass," while female voices are frequently labeled as "youthful" and "authoritative lady." These labels describe the pitch of the voice, with "teen-boy" and "youthful" indicating a higher pitch and "bass" and "authoritative lady" denoting a lower pitch. This labeling system ensures that almost every user can find a label that resonates with their voice.
Such categorization of voices into distinct labels influences user perceptions and expectations. Users tend to associate certain personality traits and characteristics with these voice types. For instance, a "teen-boy" or "youthful" label might evoke an image of someone young and energetic, while "bass" and "authoritative lady" could suggest a sense of maturity and confidence. As a result, these labels not only help users identify voices that they find appealing but also contribute to the idealization of the voice's owner based on the associated characteristics of the label, further exemplifying the impact of voice perception in online communication.}


\add{
Hosts in voice-based dating platforms offer a more authentic and efficient matchmaking process, overcoming the limitations of traditional MDAs that focus on searchable attributes, and addressing the inefficiencies found in game-based dating environments. 
The presence of a host in voice-based dating facilitates a more genuine form of matchmaking. Traditional MDAs often compel users to select partners based on searchable attributes like income or religion \cite{frost2008people}, but daters prefer to assess experiential qualities like humor or rapport. Previous research has shown that game-based dating offers a more immersive experience than typical online forums \cite{yee2011introverted}, but it lacks efficiency and consistency. In the voice-based dating site we studied, the host efficiently manages matchmaking in speed dating rooms with complex requirements and guides meaningful discussions in topic chat rooms. They also help forge user connections and foster a sense of community. However, regulating voice communication in these group environments poses more challenges compared to other communication spaces.}

\subsubsection{\add{Rationale behind Sender and Receiver Dynamics}}
\add{The \textbf{receiver}, or the listener in this case, forms first impressions of the sender's (speaker's) identity and personality based on their voice. This perception is influenced by various vocal characteristics such as timbre, accent, speech mannerisms, and tone, as detailed in Section 5.3.1 of our study.
If the receiver finds the voice timbre appealing, they are inclined to idealize the sender's identity and personalities. 
This could be because they subconsciously link the voice to that of an ideal partner. This association significantly shapes their perception and expectations during the initial stages of interaction, demonstrating the impact of voice characteristics on the formation and development of interpersonal connections in voice-based CMC environments.
The process of interpreting these vocal cues is largely subjective, depending heavily on the receiver's personal experiences and preferences. For instance, as outlined in Section 5.4.1, a receiver might choose to discontinue communication with a sender if they perceive the sender's speech as deviating from what they consider standard Mandarin. This decision is based on the receiver's interpretation and preference, demonstrating how personal biases and expectations in voice play a significant role in voice-based CMC.} 

\add{In the context of voice-based online dating, \textbf{senders} often adjust their voice and speaking style to present an idealized version of themselves, aiming to be more attractive to potential partners. 
This modification is a conscious effort by senders to align their vocal presentation with how they imagine their ideal self would speak in a dating scenario.
For example, a user like P7, who aspires to embody the persona of a ``soft girl," might find it challenging to express this identity in real-life interactions. However, in the online space, they have the opportunity to alter their voice to fit this desired persona. By softening their tone or adopting speech patterns that they associate with the ``soft girl" image, they present themselves in a way that they believe will be more appealing to others. }

\add{In the context of voice-based dating platforms, we suggest that senders might experience a lower sense of self-deception when presenting their ideal selves via voice compared to other types of self-presentation on dating sites. This could be attributed to the nature of voice communication, which, while allowing for some degree of modulation or alteration, still retains a core element of the individual's authentic self — their actual voice.
Unlike text-based or visually dominated platforms where users can heavily curate or manipulate their representation (through edited texts, photos, or entirely fabricated profiles), voice-based platforms inherently require a more genuine aspect of the individual's identity. Even when modulating tone or adopting different speech patterns to align with an idealized persona, the underlying uniqueness of one's voice remains a more authentic trait that cannot be completely disguised or altered. 
In addition, a user can easily alter their voice and speaking style to fit in their idealized self-presentation at any time.
This relative authenticity in voice communication may lead to a feeling of less self-deception for the sender. They might perceive their presentation as a more honest or genuine version of themselves, even if it is an idealized one.
}

\subsection{Voice-based MDA \add{in China} \delete{with Non-Western Values}}
\add{Individualism and collectivism cultures can influence user perceptions and behaviors towards online dating \cite{paul2022does}. 
In a country characterized by a collectivistic culture, the nature of such cultures tends to favor shared activities, such as livestreaming, over solitary ones \cite{lu2018you,lu2019feel,lu2019vicariously}. This preference may explain why community-based dating could be particularly appealing to users of Chinese dating sites. Voice chat rooms offer a unique dynamic, allowing users the choice to either distinguish themselves through speaking up or to assimilate seamlessly with the group. This flexibility can foster a sense of unity and belonging, potentially making users feel more at ease. Furthermore, topic chat rooms on platforms like Soul can significantly contribute to the development of romantic relationships by providing a space where users can connect over shared interests and commonalities. 
}

Chinese users may also have a higher privacy concerns in using MDAs. This is because MDAs have a lower acceptance to public in China compared to western countries \cite{solis2019meet}. For the MDAs that necessitate personal profile creation and image uploads pose privacy risks, they potentially leading to judgment and scrutiny from individuals in their real-life circles. \add {Chinese users, especially women, may encounter increased risks when utilizing these platforms due to the conservative social attitudes surrounding dating, relationships, and marriage in Chinese culture \cite{peng2021gender,tang2022dare}. Research and previous findings suggest that some Chinese MDA users may worry about the potential adverse consequences of disclosing their use of MDAs for romantic purposes to their family, professional networks, and community.} 
Even after marriage, some couples formed through MDAs might opt to conceal the origin of their relationship to avoid gossip and potential social stigma. These concerns underscore the challenges and hazards inherent in the intersection of modern technology and deeply entrenched cultural norms in China~\cite{blair2019dating}.
\add{By opting for voice-based dating, users can significantly mitigate the risk of being recognized through personal images, potentially enhancing their sense of security and comfort on MDAs.}

%% file: 05-Design_Implication.tex
\subsection{Implications for Design and HCI Research}





We propose several design recommendations inspired by the voice-based dating platform, which could enhance existing dating sites that rely on textual and visual information. These suggestions also offer potential avenues for further research on online dating.


\subsubsection{Use voice profile to enrich personal information} To accommodate users who are hesitant to reveal their personal images or other sensitive information, Soul promotes the use of voice profiles as an alternative method for connecting with other users. A voice profile typically consists of a brief voice recording that serves as a self-introduction. This voice profile can be incorporated into the ``swipe left, swipe right" matching mode commonly used in many MDAs to facilitate connections between users.
By listening to a voice recording, users can gain insights into another person's personality and interests, allowing them to decide whether they want to establish a connection. Integrating voice recordings into this context serves several purposes. It enhances the authenticity and genuineness of a user's profile, eliminates potential discomfort or reluctance associated with sharing sensitive personal data, and provides an alternative means for users to connect and acquaint themselves with others while protecting their privacy. 

\add{\subsubsection{Encourage diverse self-presentation types}
Online dating platforms should provide users with features that allow them to present themselves based on their dating needs and purposes, either authentically or ideally, and showcase different aspects of their personality and interests beyond their voice types. 
This might be achieved by incorporating multimedia content: Allowing users to include photos, videos, or audio recordings that showcase their interests and personality can provide a more holistic representation of themselves and encourage diverse self-presentation.}

\add{\subsubsection{Counteract homogeneous standards} 
Our research indicates that certain voice types are perceived as more attractive on voice-based dating platforms, leading users to alter their voices to fit these preferences. 
Online dating platforms should implement features that challenge users' homogeneous standards of voice by diversifying voice types of the recommended potential dates. 
This might be achieved through carefully designed algorithmic interventions. For example, platforms can develop algorithms that deliberately present users with voice types that defy stereotypical norms formed within the communities. By exposing users to a wider range of self-presentation types of voice, online dating platforms can encourage users to broaden their preferences and challenge their unconscious biases.
The platforms should also consider highlighting diverse profiles with different voice types. For example, platforms can feature diverse profiles in promotional sections or recommendations to promote the inclusivity of some rare voice types. Platforms can potentially disrupt homogeneous standards and encourage user acceptance by showcasing users who do not adhere to traditional voice standards or self-presentation types.}

\subsubsection{Add voice chat room feature} As discussed in section 4, even text-based MDAs that incorporate voice chat features may still encounter user reluctance. To address this challenge and promote increased participation in voice-based communication, one potential solution is to introduce dedicated voice chat rooms within text-based MDAs.
The inclusion of dedicated voice chat rooms provides a designated space for voice interactions, which can effectively reduce users' psychological barriers and enhance their comfort and confidence in utilizing voice communication. This is particularly advantageous for individuals who may experience anxiety or self-consciousness when engaging in text-based chat or who simply prefer a more intimate and immersive mode of communication.

\subsubsection{\add{Date in a Group Setting }\delete{Dating site as social spaces}} 
Our research indicates that group dating settings, similar to those found in interest groups or discussion rooms on other dating sites, may offer distinct advantages to users.
\delete{Our research findings emphasize the benefits of integrating social spaces,  such as interest groups or discussion rooms.} \add{These platforms provide a valuable space for users to engage and interact with others who share similar interests. Furthermore, these communities might shift the focus away from visual aspects, encouraging users to concentrate more on the shared interests and the sense of community itself. }\delete {Importantly, given that the main objective of users on dating platforms is to establish romantic relationships, these social spaces enhance the prospects of transitioning from friendships formed within dating websites to romantic partnerships. This distinct advantage sets them apart from similar social spaces that exist outside of dating platforms.}

\add{\subsubsection{Considering Other Modalities Beyond Visual}
We also advocate for expanding research into dating sites that emphasize modalities beyond the visual aspect. While this project concentrates on a voice-based platform within our demographic confines in China, there is potential novelty in designing a new platform focused on voice or other modalities, which could make significant design contributions to the field. 
}

%% file: 06-conclusion.tex
\subsection{Limitations}
Our participant recruitment was limited to a non-Western context, as a mature voice-based dating platform was only available in China at the time of our study. All of our participants are young adults in their early to mid-20s. Therefore, how users from different age groups perceive voice-based dating experiences remains unknown and we leave it for future research.

\section{Conclusion}


We conducted an interview study to explore the motivations behind using voice-based online dating apps and to understand how users interact throughout the dating process. Our findings highlight that the incorporation of voice communication within the Soul platform fosters a deeper mutual understanding among potential romantic partners, as it allows users to readily discern each other's reactions and emotions.
Beyond assisting users in finding romantic partners, the diverse dating processes and themed voice chat rooms on Soul also offer various avenues for users to forge new friendships. In summary, our research underscores the effectiveness and significance of incorporating voice communication into online dating, providing individuals with a meaningful means of connecting and getting acquainted. We anticipate that future research will further delve into the potential of novel communication methods within online dating applications.